\documentstyle[aps]{revtex}

\newcommand{\dr}[1]{\hat{\dot{#1}}}
\newcommand{\pr}[1]{#1^{\hat\prime}}
\newcommand{\ddr}[1]{\hat{\ddot{#1}}}
\newcommand{\ppr}[1]{#1^{\hat{\prime\prime}}}

\begin{document}

%%%%%%%%%%%%%%%%%%%%%%%%%%%%%%%%%%%%%%%%%%%%%%%%%%%%%%%%%%%%%%%%%%%%%%%%

\title{Gauge-invariant and coordinate-independent perturbations of
stellar collapse II: matching to the exterior}

\author{Jos\'e M. Mart\'\i n-Garc\'\i a}
\address{Universidad Alfonso X El Sabio, 
         Avenida de la Universidad 1,
         28691~Villanueva de la Ca\~nada, Madrid, Spain }
\address{Faculty of Mathematical Studies, University of Southampton,
         Southampton SO17 1BJ, UK \footnote{current address}}

\author{Carsten Gundlach} 
\address{Enrico Fermi Institute, University of Chicago, 
         5640 Ellis Avenue, Chicago, IL 60637}
\address{Faculty of Mathematical Studies, University of Southampton,
         Southampton SO17 1BJ, UK \footnote{current address}}

\date{14.12.00}

\maketitle

%%%%%%%%%%%%%%%%%%%%%%%%%%%%%%%%%%%%%%%%%%%%%%%%%%%%%%%%%%%%%%%%%%%%%%%%

\begin{abstract}
In Paper I in this series we constructed evolution equations for the
complete gauge-invariant linear perturbations of a time-dependent
spherically symmetric perfect fluid spacetime. A key application of
this formalism is the interior of a collapsing star. Here we derive
boundary conditions at the surface of the star, matching the interior
perturbations to the well-known perturbations of the vacuum
Schwarzschild spacetime outside the star.
\end{abstract}

\pacs{
04.25.Nx, 
04.30.Db,
04.40.Dg,
04.25.Dm
}

%\tableofcontents

%%%%%%%%%%%%%%%%%%%%%%%%%%%%%%%%%%%%%%%%%%%%%%%%%%%%%%%%%%%%%%%%%%%%%%%%
\section{Introduction and overview}
%%%%%%%%%%%%%%%%%%%%%%%%%%%%%%%%%%%%%%%%%%%%%%%%%%%%%%%%%%%%%%%%%%%%%%%%

In Paper I in this series \cite{fluidpert1} we constructed evolution
equations for the complete gauge-invariant linear perturbations of a
time-dependent spherically symmetric perfect fluid spacetime, with
arbitrary two-parameter equation of state $p=p(\rho,s)$. We isolated
true degrees of freedom for all perturbations. They obey a hyperbolic
system of wave and transport equations. A key application of this
formalism is the interior of a collapsing star. Here we derive
boundary conditions at the surface of the star, matching the interior
perturbations to the well-known perturbations of the vacuum
Schwarzschild spacetime outside the star. Our formalism is now
complete and ready for numerical work.

As in Paper I, we use the covariant and gauge-invariant perturbation
formalism of Gerlach and Sengupta \cite{GS1}, and our notation is
compatible with theirs. The combination of Paper I and the present
Paper II is intended to be self-contained, but we refer the reader
back to Paper I for some definitions and results. In our presentation,
we go from the general to the specific.

In Section \ref{section:junctionconditions} we discuss continuity
conditions across a hypersurface in spacetime where the stress-energy
tensor is finite but possibly discontinuous.  We argue that the
appropriate choice of continuous perturbation fields are the
perturbations of the 3-metric and extrinsic curvature of the surface
with {\it contravariant} indices. We give those quantities in terms of
metric perturbations and their first derivatives.

In Section \ref{section:spherical} we then restrict the background
spacetime to be spherically symmetric, but still allow for arbitrary
matter, and for the matching surface to be either timelike or
spacelike. We identify a complete set of continuous {\it
gauge-invariant} perturbations in the notation of Paper I, separately
for the axial and polar perturbations, decomposing all tensor
quantities into components in a frame aligned with the surface. This
simplifies and corrects results of \cite{GS2,GS3}.

In section \ref{section:fluidstar} we restrict consideration to
perfect fluid matter and a timelike matching surface, namely the
surface of the collapsing star, and state the matching conditions
across the stellar surface in terms of variables specialized to fluid
matter. The conditions for the axial perturbations are fairly simple.

In section \ref{section:polarmatching} we bring the continuity
conditions for the polar perturbations into a final form that 
shows clearly how information crosses the stellar surface in both
directions (``extraction'' and ``injection''), in a way that is
natural for a numerical evolution.

We now sketch the form our main results are going to take. In Paper I
we characterized the {\it axial} perturbations in the interior by a
tangential fluid velocity perturbation $\beta$ that obeys an
autonomous transport equation, plus a metric perturbation $\Pi$ that
obeys a wave equation with $\beta$ as a source. $\Pi$ is defined both
in the interior and exterior of the star, and obeys the same equation
in the interior and exterior. (In the exterior the matter source
$\beta$ simply vanishes.) $\beta$ requires no matching because the
stellar surface is formed by fluid worldlines and is therefore a
characteristic of the transport equation for $\beta$. The matching
condition for $\Pi$ is simply that it is continuous at the surface.

In Paper I we also characterized the $l\ge 2$ {\it polar}
perturbations by fields $\chi$, $k$ and $\psi$ that admit free Cauchy
data and evolve autonomously. By definition, these are all metric
perturbations, but we showed that while $\chi$ characterizes
gravitational waves, $k$ characterizes the sound waves, and $\psi$
characterizes a second tangential fluid velocity perturbation. (The
matter perturbations properly speaking can be reconstructed from these
primary variables by quadratures if they are needed.) Consequently,
$\chi$ and $k$ obey wave equations and $\psi$ a transport equation
(all of which are coupled). It is well known that the $l\ge 2$ polar
perturbations in the exterior are gravitational waves characterized by
a single field $Z$ that obeys a wave equation (the Zerilli equation
\cite{Zerilli}).  $\psi$ is transported along the fluid, and
therefore, like the perturbation variable $\beta$, requires no
explicit matching. The variables $\chi$, $k$ in the interior, and $Z$
in the exterior, however, obey wave equations. This means that these
three variables can and must be specified on a timelike boundary. The
matching problem for the polar perturbations therefore consists in
finding $\chi$ and $k$ just inside the surface from $Z$ and its
derivatives just outside (injection), and finding $Z$ just outside
from $\chi$, $k$, $\psi$ and their derivatives just inside
(extraction).

Matching the polar perturbations in the special cases $l=0$ and $l=1$,
which do not admit (completely) gauge-invariant perturbations and
therefore require (partial) gauge-fixing, is also discussed.

In the appendixes we compare our results to previous work, and give
some intermediate steps of the calculation.

%%%%%%%%%%%%%%%%%%%%%%%%%%%%%%%%%%%%%%%%%%%%%%%%%%%%%%%%%%%%%%%%%%%%%%%%
\section{Junction conditions on a hypersurface}
\label{section:junctionconditions}
%%%%%%%%%%%%%%%%%%%%%%%%%%%%%%%%%%%%%%%%%%%%%%%%%%%%%%%%%%%%%%%%%%%%%%%%

%%%%%%%%%%%%%%%%%%%%%%%%%%%%%%%%%%%%%%%%%%%%%%%%%%%%%%%%%%%%%%%%%%%%%%%%
\subsection{Choice of continuous perturbation objects}
%%%%%%%%%%%%%%%%%%%%%%%%%%%%%%%%%%%%%%%%%%%%%%%%%%%%%%%%%%%%%%%%%%%%%%%%

Given a spacetime $(M^4,g_{\mu\nu})$ containing matter, we want to
derive junction conditions on a hypersurface $\Sigma$ where the
stress-energy tensor is allowed to be discontinuous. Our principal
application will be the surface of a star, with fluid matter in the
interior, and vacuum in the exterior, but for now we still allow the
matching surface to be either spacelike or timelike. We assume that
the metric is at least twice differentiable on each side of the
boundary, but we do not assume that one coordinate patch covers both
sides of the boundary. Therefore we do not assume that the metric
components or their derivatives are continuous.

The condition that the stress-energy does not have a $\delta$-function
singularity on the surface (a ``surface layer'') translates into the
condition that the induced metric and extrinsic curvature of the
surface are the same on both sides of the surface \cite{Israel}. These
tensors are
\begin{eqnarray}
i_{\mu\nu} & \equiv & g_{\mu\nu} \mp n_\mu n_\nu,  \\
e_{\mu\nu} & \equiv  & n_{\mu;\alpha}{i^\alpha}_\nu = n_{\mu;\nu} \mp
n_{\mu;\alpha}n^\alpha n_\nu,
\end{eqnarray}
where $n^\mu$ is the unit vector orthogonal to the surface, with
$n_\mu n^\mu=\pm 1$.  Here and throughout this paper the upper (lower)
sign applies when $n^\mu$ is spacelike (timelike). (Therefore the
upper sign will apply when we restrict to a stellar surface later.)
The semicolon denotes a covariant derivative with respect to the
metric $g_{\mu\nu}$. The tensors $e_{\mu\nu}$ and $i_{\mu\nu}$ are
symmetric, and are intrinsic to the hypersurface in the sense that
their contraction with $n^\mu$ vanishes.

It is helpful to think of the two sides of the surface as two distinct
spacetimes, each with a boundary. The two halves match without a
surface layer if the 3-tensors $e_{\mu\nu}$ and $i_{\mu\nu}$ are
equal. This in turn means that one can locally find 3 coordinates on
each boundary so that the components of $e_{\mu\nu}$ and $i_{\mu\nu}$
are equal. To ask, for example, if $n^\mu$ is continuous is not
meaningful.

Clearly, the matching conditions for linear perturbations of the
spacetime will be the continuity of the linear perturbations of
$e^{\mu\nu}$ and $i^{\mu\nu}$, but there is a complication: because
the metric itself is perturbed, the perturbations of a tensor with its
indices in different positions are no longer the same. For the metric
itself, for example, with the notation
\begin{equation}
\Delta(g_{\mu\nu}) \equiv h_{\mu\nu},
\end{equation}
and the convention that all indices are moved with the background
metric $g_{\mu\nu}$ and its inverse $g^{\mu\nu}$, one has
\begin{equation}
\Delta(g^{\mu\nu}) = - h^{\mu\nu}.
\end{equation}
Similarly, for the perturbation of any tensor $t_{\mu\nu}$,
\begin{equation}
\Delta(t_{\mu\nu})= g_{\mu\rho} g_{\nu\lambda} \Delta(t^{\rho\lambda}) 
+ h_{\mu\lambda} {t^\lambda}_\nu + t_{\mu}{}^\lambda h_{\lambda\nu}.
\end{equation}
Which of $\Delta(e_{\mu\nu})$ and $\Delta(e^{\mu\nu})$, and of
$\Delta(i_{\mu\nu})$ and $\Delta(i^{\mu\nu})$, should be made
continuous? We are not aware of an argument in the literature, and
therefore give one here.

Let the surface be defined by the level surface $f=0$ of a scalar
field $f$. The unit normal vector is then the normalized version of
the gradient $f_{,\mu}$,
\begin{equation}
\label{normaldef}
\varphi \equiv
\left( \pm f_{,\lambda} f_{,\rho} g^{\lambda\rho} \right)^{-1/2} ,
\qquad n_\mu \equiv \varphi f_{,\mu}, 
\qquad n^\mu \equiv g^{\mu\nu} n_\nu. 
\end{equation}
The tangent vector to any curve in the surface $f=0$ gives us a vector
field $X^\mu$ that is by definition intrinsic to the surface. The
gradient field $f_{,\mu}$ is normal to any of these vectors in the
sense that $X^\mu f_{,\mu}=0$, and this fact is independent of the
metric. Similarly, a contravariant tensor $t^{\mu\nu\dots}$ is
intrinsic, independently of the metric, if and only if
$t^{\mu\nu\dots}f_{,\mu}=0$ on any of its indices. By extension we may
define $t_{\mu\nu\dots}=g_{\mu\alpha}g_{\nu\beta}\dots
t^{\alpha\beta\dots}$ to be intrinsic, but this definition of an
intrinsic covariant tensor depends on the metric. Therefore, it is
clear that the perturbations $\Delta(i^{\mu\nu})$ and
$\Delta(e^{\mu\nu})$ must be continuous (and so
$\Delta(i_{\mu\nu})$ and $\Delta(e_{\mu\nu})$ are not).

In defining linear perturbations of a spacetime by subtracting fields
at a point in the perturbed spacetime from the corresponding fields at
a point in the unperturbed spacetime, the well-known problem arises
that there is no unique or preferred map that identifies points in the
unperturbed and perturbed spacetime. Changing the map
(infinitesimally) changes the (linear) perturbations, even if the
gauge has been fixed in the background. The point identification map
is in practice provided by introducing coordinates on the perturbed
and unperturbed spacetimes, and the linear gauge freedom in the
perturbations then arises as the absence of a unique or preferred
coordinate system on the perturbed spacetime, even if the background
has a preferred coordinate system adapted to its symmetries. The gauge
freedom in the linear perturbations arises over and above the gauge
freedom in the background.

The surface of a star has a coordinate-independent significance, and
it is therefore natural to identify the perturbed with the unperturbed
surface. In fact, subtracting a background spacetime from the
perturbed spacetime makes sense {\it only} if the point identification
map maps the perturbed surface to the unperturbed surfaces: otherwise
we would have to subtract, for example, a point in the fluid interior
of the perturbed star from a point in the vacuum exterior of the
background star. The resulting density ``perturbation'' would then be
the background density, which in general is not small on the surface.

To avoid this problem, we choose a perturbation gauge in which the
perturbed surface coincides with the background surface, obtain the
matching conditions, and then transform back to the perturbation gauge
in which we actually want to work. To formalize this, let us call
$\Delta X$ the perturbation of a tensor $X$ in a completely arbitrary
gauge. By Lie-dragging $\Delta X$ along the vector field
\begin{equation}
\xi^\mu \equiv \pm \varphi \Delta f n^\mu.
\end{equation}
we obtain the same perturbation in a new, auxiliary, gauge:
\begin{equation}
\bar\Delta X\equiv \Delta X - {\cal L}_{\xi} X
\end{equation}
In this gauge the surface has not moved, because we see that
$\bar\Delta f=0$. We call this the surface gauge. (This gauge is of
course not unique. Rather out of the four gauge degrees of freedom in
a completely arbitrary gauge, the three parallel to the surface remain
free.) In the following, we carry out some intermediate steps in
surface gauge, and indicate this by using an overbar. As long as our
final result contains only gauge-invariant fields, it does not matter
in which (partial) gauge we have obtained it.

%%%%%%%%%%%%%%%%%%%%%%%%%%%%%%%%%%%%%%%%%%%%%%%%%%%%%%%%%%%%%%%%%%%%%%%%
\subsection{Continuous perturbation objects in terms of metric perturbations}
%%%%%%%%%%%%%%%%%%%%%%%%%%%%%%%%%%%%%%%%%%%%%%%%%%%%%%%%%%%%%%%%%%%%%%%%

In surface gauge, with $\bar \Delta f=0$, the perturbation of the
normal vector $n_{\mu}$ is
\begin{equation}
\bar\Delta(n_\mu) 
= \pm {1\over 2} n^\lambda n^\rho \bar h_{\lambda\rho} n_\mu,
\end{equation}
so that
\begin{equation}
\bar\Delta(n^\mu) 
= \pm {1\over 2} n^\lambda n^\rho \bar h_{\lambda\rho} n^\mu -
\bar h^{\mu\nu} n_{\nu}.
\end{equation}
Note that $\bar\Delta(n_\mu)$ is proportional to $n_\mu$, while
$\bar\Delta(n^\mu)$ is not in general proportional to $n^\mu$. This is
another reflection of the fact that the natural intrinsic objects are
contravariant: if their contraction with $n_\mu$ vanishes, so does
their contraction with $n_\mu + \bar\Delta(n_\mu)$.  For the perturbation
of a covariant derivative we use the formula
\begin{equation}
\Delta({X^\mu}_{;\nu}) = \left[\Delta(X^\mu)\right]_{;\nu} + \Delta
({\Gamma^\mu}_{\lambda\nu}) X^\lambda
\end{equation}
and its obvious extensions to other tensors, where the ``perturbation of
the Christoffel symbol'' is a tensor:
\begin{equation}
\Delta
{\Gamma^\mu}_{\lambda\nu} \equiv {1\over2} g^{\mu\rho}\left(
h_{\rho\lambda;\nu} + h_{\rho\nu;\lambda} -
h_{\lambda\nu;\rho}\right).
\end{equation}
The continuous perturbations in surface gauge are
\begin{eqnarray}
\label{deltaiup}
\bar\Delta(i^{\mu\nu}) & = & 
- i^{\mu\alpha}i^{\nu\beta} \bar h_{\alpha\beta}, \\
\label{deltaeup}
\bar\Delta(e^{\mu\nu}) & = & 
\left [\pm {1\over 2}  e^{\mu\nu}n^\alpha n^\beta 
-(i^{\mu\alpha} e^{\nu\beta} + i^{\nu\beta} e^{\mu\alpha}) 
\right] \bar h_{\alpha\beta}
- i^{\mu\alpha}i^{\nu\beta} n^\lambda
{1\over 2} (\bar h_{\lambda\alpha;\beta} +
\bar h_{\lambda\beta;\alpha} -\bar h_{\alpha\beta;\lambda})  .
\end{eqnarray}
We can replace continuity of $\bar\Delta(e^{\mu\nu})$ by continuity of
the shorter expression
\begin{equation}
\label{deltaeupvariant}
\bar\Delta(e^{\mu\nu})  
- {e_\alpha}^\mu \bar\Delta(i^{\alpha\nu})  
- {e_\alpha}^\nu\bar\Delta(i^{\mu\alpha})=
\pm {1\over 2}  e^{\mu\nu}n^\alpha n^\beta \bar h_{\alpha\beta}
- i^{\mu\alpha}i^{\nu\beta} n^\lambda
{1\over 2} (\bar h_{\lambda\alpha;\beta} +
\bar h_{\lambda\beta;\alpha} -\bar h_{\alpha\beta;\lambda}).
\end{equation}
$\bar\Delta(i_{\mu\nu})$ and $\bar\Delta(e_{\mu\nu})$ are not continuous, but
one finds by explicit calculation that
\begin{equation}
i^{\mu\alpha}i^{\nu\beta} \bar\Delta(i_{\alpha\beta}), \qquad
i^{\mu\alpha}i^{\nu\beta} \bar\Delta(e_{\alpha\beta})
\end{equation}
are continuous. (Gerlach and Sengupta \cite{GS2,GS3} take the
continuity of these expressions, or rather their equivalent in a
general gauge, as their starting point.)

%%%%%%%%%%%%%%%%%%%%%%%%%%%%%%%%%%%%%%%%%%%%%%%%%%%%%%%%%%%%%%%%%%%%%%%%
\section{Perturbations of spherical symmetry}
\label{section:spherical}
%%%%%%%%%%%%%%%%%%%%%%%%%%%%%%%%%%%%%%%%%%%%%%%%%%%%%%%%%%%%%%%%%%%%%%%%

%%%%%%%%%%%%%%%%%%%%%%%%%%%%%%%%%%%%%%%%%%%%%%%%%%%%%%%%%%%%%%%%%%%%%%%%
\subsection{Spherical background}
%%%%%%%%%%%%%%%%%%%%%%%%%%%%%%%%%%%%%%%%%%%%%%%%%%%%%%%%%%%%%%%%%%%%%%%%

We now restrict ourselves to a spherically symmetric background, but
still allow the matching surface to be either timelike (upper sign in
all following equations) or spacelike (lower sign). As in Paper I, we
write the spherical background metric in a 2+2 covariant decomposition
as
\begin{equation}
g_{\mu\nu} = {\rm diag} \left(g_{AB},r^2 \gamma_{ab}\right),
\end{equation}
where $\gamma_{ab}$ is the unit metric on the two-sphere,
and the background stress energy tensor as
\begin{equation}
t_{\mu\nu} = {\rm diag} \left(t_{AB},Q r^2 \gamma_{ab}\right).
\end{equation}
Let $n^A$ be the vector field normal to the matching surface, with
length squared $\pm 1$. Let $u^A$ be the vector field tangential to
the surface, with length squared $\mp 1$. The 1+1 metric $g_{AB}$ can
be written in terms of this orthonormal basis as
\begin{equation}
g_{AB}=\pm(-u_Au_B+n_An_B).
\end{equation}
We also use the notation $g_{AB|C}\equiv 0$ for the
covariant derivative in two dimensions, $v_A\equiv r^{-1} r_{,A}$, and
the frame unit derivatives $\dot f\equiv u^A f_{,A}$ and $f'\equiv n^A
f_{,A}$.

The induced metric and extrinsic curvature of the matching surface of
the background spacetime are
\begin{eqnarray}
i_{\mu\nu} & = & {\rm diag}(\mp u_A u_B, r^2 \gamma_{ab}), \\
e_{\mu\nu} & = & {\rm diag}(\mp \nu \, u_A u_B , W r^2 \gamma_{ab}
).
\end{eqnarray}
Therefore the scalars $r$, $\nu\equiv{n^A}_{|A}$ and $W\equiv n^A
v_A=r'/r$ are continuous. Note that although $\nu$ contains a
derivative of $n^A$, it depends only on the surface itself. As $r$ is
continuous everywhere on the surface, its unit derivative $\dot r$
along the surface, and therefore $U \equiv u^A v_A$, is
continuous. From the continuity of $\dot r$ and $r'$ follows the
continuity of $r_{,A} r^{,A}$ and hence of the Hawking mass $m$. From
the Einstein equations we have that $n^A n^Bt_{AB}$ and $u^A n^B
t_{AB}$ are continuous. The list of continuous quantities can be
extended by taking dot-derivatives of continuous quantities. The
quantities that are {\it not} required to be continuous include $u^A
u^B t_{AB}$, $Q$, the Gauss curvature ${\cal R}$ of $g_{AB}$, and
$\mu\equiv {u^A}_{|A}$.  Note that at the surface of a perfect fluid
star (where $u_A$ coincides with the fluid 4-velocity) $u^A n^B
t_{AB}$ and $Q$ vanish identically.

%%%%%%%%%%%%%%%%%%%%%%%%%%%%%%%%%%%%%%%%%%%%%%%%%%%%%%%%%%%%%%%%%%%%%%%%
\subsection{Axial perturbations}
%%%%%%%%%%%%%%%%%%%%%%%%%%%%%%%%%%%%%%%%%%%%%%%%%%%%%%%%%%%%%%%%%%%%%%%%

In the following we use the notation of \cite{fluidpert1} for the
gauge-dependent and the gauge-invariant matter and metric
perturbations, without repeating all the definitions. That notation
uses the general metric and stress-energy perturbations of Gerlach and
Sengupta \cite{GS1}, adding only specific notation for perfect fluid
matter. Our notation here is also compatible with the notation of
Gerlach and Sengupta in their papers on the matching conditions
\cite{GS2,GS3}. Note that for the axial perturbations we do not need
to use surface gauge, as $\Delta f$ is a polar perturbation.

The general junction conditions for axial perturbations in the
framework of Gerlach and Sengupta are continuity of
\begin{eqnarray}
l\ge 1: &\quad& u^A h_A^{\text{axial}}, \\
l\ge 2: &\quad& h,
\end{eqnarray}
from continuity of the induced metric (\ref{deltaiup}), and
\begin{eqnarray}
l\ge 1: &\quad& u^A n^B \left( h_{A|B}^{\text{axial}} 
                             - h_{B|A}^{\text{axial}} 
                             + 2 v_A h_B^{\text{axial}} 
                      \right), \\
l\ge 2: &\quad& n^A \left(h_A^{\text{axial}} - h_{,A}\right),
\end{eqnarray}
from continuity of the extrinsic curvature (\ref{deltaeupvariant}). 

In the axial sector we have four gauge-dependent continuity conditions
(for $l\ge 2$) and one gauge freedom. Therefore we find three
gauge-independent continuous objects:
\begin{eqnarray}
&& l\ge 1: \qquad \Pi\equiv \epsilon^{AB}(r^{-2}k_A)_{|B}, \\
\label{kcontinuity}
&& l\ge 2: \qquad  n^A k_A, \quad u^A k_A.
\end{eqnarray}
$\Pi$ is of particular interest as it obeys a wave equation with
source terms given purely by matter perturbations. The Einstein
equations allow us to reconstruct $k_A$ later from $\Pi$ and the
matter perturbation $L_A$, as
\begin{equation}
\label{Pikconstraint}
l\ge 1: \qquad
(l-1)(l+2) k_A=16\pi r^2L_A-\epsilon_{AB}(r^4\Pi)^{|B}.
\end{equation}
Here $\epsilon_{AB}$ is the totally antisymmetric covariant tensor
with respect to $g_{AB}$. It can be given in terms of the basis as
\begin{equation}
\epsilon_{AB}=\pm(n_Au_B-u_An_B)
\end{equation}
We can use this to translate the continuity conditions on $k_A$ into
conditions on $\Pi$, the only dynamical variable we work with. As
$\Pi$ is continuous at all times at the matching surface, its
derivative along the unit vector $u^A$ in the surface, $\dot\Pi$, must
also be continuous. Multiplying (\ref{Pikconstraint}) by $n^A$ and using
(\ref{kcontinuity}) when $l\ge 2$, we find continuity of
\begin{equation}
\label{nALAcontinuity}
l\ge 1: \qquad  n^A L_A.
\end{equation}
This condition must be obeyed automatically if we have chosen the
matching surface consistently with the matter equations of
motion. Multiplying (\ref{Pikconstraint}) by $u^A$ we find that
\begin{equation}
\label{uALAcontinuity}
l\ge 1: \qquad \Pi'-16\pi r^{-2} u^A L_A
\end{equation}
is continuous, where we have used continuity of $r'$. This is a
condition that needs to be imposed on the initial data for $\Pi$ and
the matter. It is compatible with the matter conservation
equation. The continuity of $\Pi$ itself needs to be imposed during
the time evolution as a dynamical boundary condition.

For $l=1$ the variable $\Pi$ is still gauge-invariant, but it no
longer obeys a wave equation. Instead it is determined by the matter
perturbations \cite{fluidpert1} through
\begin{equation}
l=1: \qquad r^4 \Pi = 16\pi T, \qquad \text{where} \qquad 
r^2 L_A = \epsilon_{AB}T^{|B} .
\end{equation}
The continuity of $\Pi$ implies the continuity of $T$, and
(\ref{nALAcontinuity}) and (\ref{uALAcontinuity}) still hold. For
$l=0$, there are no gauge-invariant axial perturbations. Our 
results coincide exactly with those of Gerlach and Sengupta in
\cite{GS2}.

%%%%%%%%%%%%%%%%%%%%%%%%%%%%%%%%%%%%%%%%%%%%%%%%%%%%%%%%%%%%%%%%%%%%%%%%
\subsection{Polar perturbations}
%%%%%%%%%%%%%%%%%%%%%%%%%%%%%%%%%%%%%%%%%%%%%%%%%%%%%%%%%%%%%%%%%%%%%%%%

The general junction conditions for polar perturbations in surface
gauge are continuity of
\begin{eqnarray}
l\ge 0: &\qquad& \bar i_1 = u^A u^B \bar h_{AB}, \\
        &\qquad& \bar i_2 = \bar K, \\
l\ge 1: &\qquad& \bar i_3 = u^A \bar h_A^{\text{polar}}, \\
l\ge 2: &\qquad& \bar i_4 = \bar G .
\end{eqnarray}
from continuity of the induced metric, and
\begin{eqnarray}
l\ge 0: \qquad
\bar e_1 &=& u^A u^B n^C \left(\bar h_{CA|B} 
+ \bar h_{CB|A} - \bar h_{AB|C} \right) 
+ {n^C}_{|C} n^A n^B
\bar h_{AB}, \\
\bar e_2 &=& n^A \left(2\bar h_{AB} v^B - \bar K_{,A} - 2 v_A \bar K
- \frac{l(l+1)}{r^2} h^{\text{polar}}_A \right) 
\mp n^C v_C n^A n^B \bar h_{AB} , \\
l\ge 1: \qquad
\bar e_3 &=& u^A n^B \left(\bar h_{AB} - \bar h_{A|B}^{\text{polar}} +
\bar h_{B|A}^{\text{polar}}
- 2 v_A \bar h_B^{\text{polar}}\right), \\
l\ge 2: \qquad
\bar e_4 &=&n^A\left(\bar h_A^{\text{polar}} 
- {1\over 2} r^2 \bar G_{,A} - r^2 v_A \bar G\right)
\end{eqnarray}
from continuity of the extrinsic curvature. Note that these are all
scalars, and from their continuity at all time follows that of their
derivative along the surface, e.g. of $u^A K_{,A}\equiv \dot K$. 

Now we go back to the general gauge in which $\Delta f$ is
arbitrary. The continuous fields $i_1$ to $e_4$ in general gauge are
$\bar i_1$ to $\bar e_4$ plus terms proportional to $\Delta f$ and its
derivatives. Next we find those linear combinations that are
gauge-invariant for $l\ge 2$. As a rule of thumb, this is done by
eliminating $G$ and both components of the vector $p_A \equiv
h_A^{\text{polar}} - (1/2)r^2 G_{|A}$. From this argument we would
expect ``$8-3=5$'' continuous gauge-invariant fields. But we note that
\begin{equation}
N\equiv \varphi \Delta f - p^A n_A = \varphi \left(\Delta f - p^A
f_{,A} \right) ,
\end{equation}
is the gauge-invariant perturbation of the scalar $f$ times the
background quantity $\varphi$. Therefore $N$ is gauge-invariant. In
surface gauge it reduces to $-n^A p_A=-e_4$, which is
continuous. Therefore we have 6 rather than 5 continuous
gauge-invariant perturbations. ($N$ is the same variable as the $N$ of
Gerlach and Sengupta \cite{GS3}.)

$N$ describes the deformation of the surface: while it is formally
gauge-invariant in the bulk, it has a physical meaning only on the
surface. To understand its significance better, we note that the
perturbed $f$ vanishes on the perturbed surface, so that
\begin{equation}
[f+\Delta f](x^\mu + \Delta x^\mu)=\Delta f+f_{,\mu}\Delta
x^\mu+O(\Delta^2) = 0,
\end{equation}
so that, in any gauge,
\begin{equation}
\label{N_coords}
N=-\varphi\Delta x^A f_{,A} - p^A n_A = -n_A (\Delta x^A+p^A).
\end{equation}
The ``surface displacement'' $\Delta x^\mu$ is a gauge-dependent
perturbation that transforms as vector field. Regge-Wheeler (RW) gauge
is by definition the gauge where $p^A=0$, so we can characterize $-N$
as the normal proper displacement of the surface in RW gauge. Note
that this is a gauge-dependent statement. In comoving gauge, for
example, the surface is by definition not displaced.

The six continuous fields are
\begin{eqnarray}
l\ge 0: \quad
\label{I_1}
I_1 &=& u^A u^B k_{AB} + 2 \nu N , \\
I_2 &=& k \mp 2 W N , \\
E_1 &=& \mp 2 {\cal R} N + \nu n^A n^B k_{AB} + u^Au^Bn^C k_{AB|C}
   - 2 \left(u^An^Bk_{AB}-\dot{N}\right)\dot{}
   - 2\nu^2 N , \\
E_2 &=& \pm W n^A n^B k_{AB} - k' \mp 8\pi u^A u^B t_{AB} N
   \pm \left( U^2 - 3 W^2 + \frac{1\pm l(l+1)}{r^2} \right) N
   \mp 2 U \left( u^A n^B k_{AB} - \dot{N} \right) , \\
l\ge 1: \quad
E_3 &=& u^An^B k_{AB} - 2\dot{N} + 2UN , \\
l\ge 2: \quad
\label{E_4}
E_4 &=& N .
\end{eqnarray}
These fields are continuous in any gauge and for the
values of $l$ indicated. They are gauge-invariant for
$l\ge2$, but only partially gauge-invariant for $l=1$, and not
gauge-invariant for $l=0$.

%%%%%%%%%%%%%%%%%%%%%%%%%%%%%%%%%%%%%%%%%%%%%%%%%%%%%%%%%%%%%%%%%%%%%%%%
\section{Perfect fluid matter}
\label{section:fluidstar}
%%%%%%%%%%%%%%%%%%%%%%%%%%%%%%%%%%%%%%%%%%%%%%%%%%%%%%%%%%%%%%%%%%%%%%%%

We now specialize to perfect fluid matter, and to the case where $n^A$
is spacelike. We therefore take the upper sign in the equations
above. In the stress-energy tensor, we have
\begin{equation}
u^A u^B t_{AB} = \rho, \quad
n^A n^B t_{AB} = p, \quad 
u^A n^B t_{AB} = 0, \quad
Q=p. 
\end{equation}
The surface on the star is defined by $p=0$, and $p$ is therefore
continuous, but $\rho$ can be discontinuous. The fluid four-velocity
is tangent to the surface, and therefore we use the notation $u^A$ for
both the unit tangent vector to the surface, and for the fluid
four-velocity inside, and we use $n^A$ both for the vector that is
normal to the surface, and normal to the fluid four-velocity in the
interior of the star.  As in paper I, we choose $n^A$ to point outside
and $u^A$ to the future.  $f$ must then increase with radius to obtain
$f'>0$ in accordance with the definition (\ref{normaldef}).

%%%%%%%%%%%%%%%%%%%%%%%%%%%%%%%%%%%%%%%%%%%%%%%%%%%%%%%%%%%%%%%%%%%%%%%%
\subsection{Axial perturbations}
%%%%%%%%%%%%%%%%%%%%%%%%%%%%%%%%%%%%%%%%%%%%%%%%%%%%%%%%%%%%%%%%%%%%%%%%

In the interior, the dynamical degrees of freedom are a matter
velocity perturbation $\beta$ and a metric perturbation $\Pi$. In
order to work with fields that are $O(1)$ at the origin, we rescale
them as
\begin{equation}
\beta\equiv r^{l+1} \bar\beta,
\quad \Pi\equiv r^{l-2} \bar \Pi,
\end{equation}
$\bar\beta$ obeys an autonomous transport equation (it is, of course,
transported along with the background fluid), while $\bar\Pi$ obeys a
wave equation with $\bar\beta$ as a source term (Eqs. (65) and (71) of
Paper I). In the exterior, the matter perturbation $\bar\beta$ is not
defined, but the metric perturbation $\bar\Pi$ is, and it obeys the
same field equation just without its $\bar\beta$ source term. We can
therefore work with the same equations of motion in both the interior
and exterior, with $\bar\beta$ defined to vanish identically in the
exterior. $\bar\beta$ parameterizes tangential fluid motion, and
therefore there is no reason why it should vanish just inside the
surface. It will therefore be discontinuous at the stellar
surface. This discontinuity is consistent with the equation of motion,
as $\bar\beta$ is transported along fluid worldlines, and so could be
discontinuous between any two fluid world lines.

Continuity of $\Pi$ and $\dot\Pi$ is equivalent to continuity of
$\bar\Pi$ and $\dot{\bar\Pi}$, as $r$ and $\dot r$ are continuous.
For perfect fluid matter, the stress-energy perturbation $L_A$ is
given by
\begin{equation}
L_A =  \beta (\rho + p) u_A, 
\end{equation}
and so $L_A n^A$ vanishes identically. It is therefore automatically
continuous, as suggested above. Continuity of (\ref{uALAcontinuity})
is equivalent
to continuity of
\begin{equation}
\label{junction_Piprime}
{\bar\Pi'}+16\pi r\rho\bar\beta
\end{equation}
in terms of the rescaled variables,
where we have used continuity of $\bar\Pi$ and $r'$. 

Matching for the axial perturbations is straightforward.  The
continuity of (\ref{junction_Piprime}) is a constraint on the initial
data $\bar\Pi$, $\dot{\bar\Pi}$ and $\bar\beta$. It is conserved by
the evolution equations. (Note that $\bar\Pi'$ is generally
discontinuous if $\rho\ne0$ on the surface.) During the evolution, one
has to impose the continuity of $\bar\Pi$. The case $l=1$ requires no
matching at all, as $\Pi$ is not a dynamical variable.

%%%%%%%%%%%%%%%%%%%%%%%%%%%%%%%%%%%%%%%%%%%%%%%%%%%%%%%%%%%%%%%%%%%%%%%%
\subsection{Polar perturbations}
\label{subsection:polar_perturbations}
%%%%%%%%%%%%%%%%%%%%%%%%%%%%%%%%%%%%%%%%%%%%%%%%%%%%%%%%%%%%%%%%%%%%%%%%

The variable $N$ is not defined in the bulk but only on the
boundary. As we want matching conditions for the bulk variables, we
must eliminate $N$ from the matching conditions. Here we do this for
perfect fluid matter.

The stellar surface is defined by $p=0$, and so it is natural to
identify $f$ with $-p$, with the minus sign chosen so that $f$
increases with radius. It does not matter that $p$ is not defined in
the exterior, as we shall only need to calculate $N$ just inside the
surface. So $N$ is related to the pressure perturbation, which in turn
is determined by the density perturbation $\omega$ and entropy
perturbation $\sigma$ through the equation of state. In appendix
\ref{appendix:N}, we derive the following result:
\begin{equation}
\label{Nomega}
l\ge 0:\qquad N = - \left. \frac{c_s^2 \omega + C
\sigma}{\nu}\right|_{\text{just inside the surface}}.
\end{equation}
Note that if $c_s^2$ vanishes on the surface (as it does for example
for polytropic equations of state) $\omega$ diverges while $N$ is finite.

A second approach to calculating $N$ is to relate the time derivative
of the radial surface displacement $N$ to the radial fluid velocity
perturbation $\gamma$. In appendix \ref{appendix:N}, we show that this
leads to the expression
\begin{equation}
\label{Ndot}
l\ge 0:\qquad \dot{N} - \mu N = - \left(\gamma +
\frac{\psi}{2}\right)_{\text{just inside the surface}}.
\end{equation}
We can use this equation as a better behaved alternative to
(\ref{Nomega}) by integrating it as an ODE in time along the
surface. 

If one combines the dot-derivative of (\ref{Nomega}) with
(\ref{Ndot}), one does not obtain an identity, but a new boundary
condition. Using the equation for $\dot\sigma$, and the fact that
$c_s^2$ and $C$ are constant on the surface, it can be written as
\begin{equation}
\label{Nboundary}
l\ge 0: \qquad c_s^2 \dot\omega = \left(\mu+{\dot\nu\over\nu}\right)
(c_s^2\omega+C\sigma) +\left(\gamma+{\psi\over2}\right)(\nu+Cs')
\quad\hbox{on the surface}.
\end{equation}
This equation relates $\dot\omega$ to $\gamma$, while the matter
equation of motion in the bulk relates $\dot\omega$ to $\gamma'$
(Eq. (99) of Paper I). Using the bulk equation for $\dot\omega$, we
obtain a relation between $\gamma$ and $\gamma'$. This new equation
can be used as a dynamical boundary condition governing the reflection
of sound waves at the boundary. But its origin is the purely kinematic
boundary condition that the Lagrangian pressure perturbation
$\bar\Delta p$ vanishes at the surface.

We can now address the continuity conditions for the bulk variables.
We use the Einstein equations to replace $\cal R$. We also express the
tensor $k_{AB}$ in terms of its components in the fluid
frame. Finally, we assume that the stellar surface is timelike, and so
pick the upper sign. The continuous perturbation fields then become
\begin{eqnarray}
l\ge 0: \quad
I_1 &=& \chi+k-2\eta + 2 \nu N , \\
I_2 &=& k - 2 W N , \\
E_1 &=& \left[8\pi(\rho+p)-{4m\over r^3}\right] N 
+ \nu (\chi+k) + (\chi+k-2\eta)'+2\mu\psi
   + 2\dot\psi + 2\ddot N
   - 2\nu^2 N , \\
E_2 &=& W(\chi+k) - k' - 8\pi\rho N
   + \left( U^2 - 3 W^2 + \frac{1\pm l(l+1)}{r^2} \right) N
   + 2 U \left(\psi+ \dot{N} \right) , \\
l\ge 1: \quad
E_3 &=& -\psi - 2\dot{N} + 2UN , \\
l\ge 2: \quad
E_4 &=& N ,
\end{eqnarray}
where $N$ is given in terms of the matter perturbations by either
(\ref{Nomega}) or (\ref{Ndot}).  Note again that while these fields
are continuous in any gauge, for the values of $l$ indicated, they are
completely gauge-invariant only for $l\ge2$. We now consider the three
cases $l=0$, $l=1$ and $l\ge 2$ separately.

%%%%%%%%%%%%%%%%%%%%%%%%%%%%%%%%%%%%%%%%%%%%%%%%%%%%%%%%%%%%%%%%%%%%%%%%
\subsubsection{$l\ge 2$}
%%%%%%%%%%%%%%%%%%%%%%%%%%%%%%%%%%%%%%%%%%%%%%%%%%%%%%%%%%%%%%%%%%%%%%%%

For perfect fluid matter, $\eta=0$ for $l\ge2$ by virtue of one of the
perturbed Einstein equations. Taking linear combinations of the six
continuous fields $I_1$ to $E_4$ and their dot-derivatives, we obtain
a simplified set of continuous gauge-invariant fields:
\begin{eqnarray}
l\ge 2: \quad C_1 &=& N, \\
C_2 &=& k, \\
C_3 &=& \chi, \\
C_4 &=& \psi, \\
C_5 &=& k' + 8\pi\rho N, \\
C_6 &=& \chi' + 2 \mu \psi .
\end{eqnarray}
Note that the coefficients $\rho$ and $\mu$ can be discontinuous on
the stellar surface. Note also that we do not need the value of $N$ in
the exterior, as its coefficient $8\pi\rho$ vanishes there.

%%%%%%%%%%%%%%%%%%%%%%%%%%%%%%%%%%%%%%%%%%%%%%%%%%%%%%%%%%%%%%%%%%%%%%%%
\subsubsection{$l=1$}
%%%%%%%%%%%%%%%%%%%%%%%%%%%%%%%%%%%%%%%%%%%%%%%%%%%%%%%%%%%%%%%%%%%%%%%%

For $l=1$, $\eta$ does not vanish, and $E_4$ is no longer
continuous. As our variables are now only partially gauge-invariant,
we make the partial gauge choice $k=0$. In this partial gauge, the
following five fields are continuous, and gauge-invariant under the
remaining gauge freedom:
\begin{eqnarray}
\label{l=1contfirst}
l=1,k=0:\quad \tilde C_1 &=& N, \\
\tilde C_2 &=& \chi-2\eta, \\
\tilde C_3 &=& \psi, \\
\tilde C_4 &=& W\chi-8\pi\rho N, \\
\label{l=1contlast}
\tilde C_5 &=& (\nu+W)\chi + (\chi-2\eta)' + 2\mu\psi.
\end{eqnarray}
Note that $\chi$ is not continuous for $l=1$. 

%%%%%%%%%%%%%%%%%%%%%%%%%%%%%%%%%%%%%%%%%%%%%%%%%%%%%%%%%%%%%%%%%%%%%%%%
\subsubsection{$l=0$}
%%%%%%%%%%%%%%%%%%%%%%%%%%%%%%%%%%%%%%%%%%%%%%%%%%%%%%%%%%%%%%%%%%%%%%%%

For $l=0$, both $E_3$ and $E_4$ are no longer continuous. Our
variables are now not gauge-invariant at all. We begin by fixing the
gauge partially, again by setting $k=0$. The four continuous (but
gauge-dependent) fields in this partial gauge can be written as
\begin{eqnarray}
l=0,k=0:\quad \hat C_1 &=& N, \\
\hat C_2 &=& \chi-2\eta, \\
\hat C_3 &=& W\chi-8\pi\rho N+2U\psi, \\
\hat C_4 &=& (\nu+W)\chi + (\chi-2\eta)' + 2(\mu+U)\psi +2\dot\psi.
\end{eqnarray}
Note that $\chi$ and $\psi$ are not continuous for $l=0$. 

%%%%%%%%%%%%%%%%%%%%%%%%%%%%%%%%%%%%%%%%%%%%%%%%%%%%%%%%%%%%%%%%%%%%%%%%
\section{Matching conditions for the polar perturbations}
\label{section:polarmatching}
%%%%%%%%%%%%%%%%%%%%%%%%%%%%%%%%%%%%%%%%%%%%%%%%%%%%%%%%%%%%%%%%%%%%%%%%

The physical $l\ge 2$ polar perturbations comprise fluid convection,
sound waves and gravitational waves in the interior, characterized by
variables $\psi$, $k$ and $\chi$ respectively, but all coupled
together. In the exterior there are only gravitational waves, and
these can be characterized by a single variable $Z$ first found by
Zerilli. The matching is non-trivial because at the surface $Z$ does
not simply coincide with $\chi$.

To achieve clarity, we formulate the matching conditions as two sets
of boundary conditions. $\chi$, $\psi$ and $k$ are considered as
evolving on a spacetime with a timelike boundary, on which one can
freely specify certain boundary conditions. The most natural ones are
Dirichlet boundary conditions for $\chi$ and $k$, the two variables
that obey wave equations. Similarly, $Z$ obeys a wave equation on
another spacetime with timelike boundary, and one can freely specify a
Dirichlet boundary condition for it. In this view the matching problem
consists in finding Dirichlet boundary conditions for $Z$ given
$\chi$, $\psi$ and $k$ and their first derivatives in the interior
(extraction), and Dirichlet boundary conditions for $\chi$ and $k$
given $Z$ and its first derivatives in the exterior (injection). 

%%%%%%%%%%%%%%%%%%%%%%%%%%%%%%%%%%%%%%%%%%%%%%%%%%%%%%%%%%%%%%%%%%%%%%%%
\subsection{Vacuum exterior}
%%%%%%%%%%%%%%%%%%%%%%%%%%%%%%%%%%%%%%%%%%%%%%%%%%%%%%%%%%%%%%%%%%%%%%%%

The exterior spacetime in spherical symmetry must be the Schwarzschild
spacetime, with metric
\begin{equation}
ds^2 = -a^{-2} \, dt^2 + a^2\, dr^2 + r^2 \, d\Omega^2,
\end{equation}
where $a^2$ is shorthand for $(1-2m/r)^{-1}$. Here $r$ and $t$ are the
Schwarzschild coordinates, and $m$ is the mass of the star. $r$ is
identical with the area radius $r$, and $m$ is identical with the
Hawking mass $m$, which is constant in the exterior.  The matching
conditions in the background spacetime are continuity of $r$ and $m$
at the stellar surface. As in Paper I, we use a hat to distinguish the
radial frame from the fluid frame. On the Schwarzschild background, in
the radial frame, we have
\begin{equation}
\label{radial_frame}
\hat U=\hat \mu=0, \quad \hat W=(ar)^{-1}, 
\quad \hat \nu = 
(a^{-1})_{,r}=mr^{-2} a. 
\end{equation}
The frame derivative $\dr{}$ of all these quantities also vanishes.
The frame derivatives expressed in Schwarzschild coordinates are
$\pr{f}=a^{-1}f_{,r}$ and $\dr{f}=af_{,t}$. It is useful to introduce
the tortoise radius $r_*$ by $dr_*=a^2(r)dr$, so that
$\pr{f}=af_{,r_*}$.

Zerilli \cite{Zerilli} first found that the $l\ge 2$ polar perturbations of
Schwarzschild obey a single wave equation for a single variable, with
all other perturbation variables obtained by quadratures. A
gauge-invariant single variable obeying the same wave equation (the
Zerilli equation) was constructed by Moncrief \cite{Moncrief}. In
appendix \ref{appendix:Zerilli} we rederive the Moncrief variable and
the Zerilli equation in our framework. The
final result is that the variable
\begin{equation}
\label{Zerilli_def}
Z\equiv A(r)\hat\chi+B(r)k+C(r)\pr{k}
\end{equation}
with 
\begin{equation}
\label{Zerilli_coeffs}
A=\frac{2r}{a^2(l^2+l+1)-3}, \quad
B=\frac{r[a^2(l^2+l+1)-1]}{a^2(l^2+l+1)-3}, \quad
C=\frac{- 2ar^2}{a^2(l^2+l+1)-3},
\end{equation}
obeys the wave equation
\begin{equation}
\label{Zerilli_eqn}
Z^{|A}{}_{|A} - V(r) Z = -\ddr{Z} + \ppr{Z}+ \hat{\nu} \pr{Z} - V(r) Z = 0,
\end{equation}
with the potential
\begin{equation}
V(r)= \frac{l(l+1)}{r^2}
- \frac{6m}{r^3} \frac{r^2\lambda(\lambda+2)+3m(r-m)}{(r\lambda+3m)^2},
\end{equation}
where $\lambda\equiv(l+2)(l-1)/2$.  Note that the Zerilli equation can
also be written as $-Z_{,tt} + Z_{,r_*r_*} - a^{-2} V(r) Z = 0$
using the tortoise coordinate. The initial data $Z$, $\dr{Z}$ can be set
freely on a Cauchy surface, and $Z$ evolves autonomously. The other
metric perturbations are given algebraically in terms of derivatives
of $Z$, and obey the Einstein equations automatically.

For $l=0$ and $l=1$ all exterior perturbations are pure gauge. We
review this in appendix \ref{appendix:l=0,1exterior}. 

%%%%%%%%%%%%%%%%%%%%%%%%%%%%%%%%%%%%%%%%%%%%%%%%%%%%%%%%%%%%%%%%%%%%%%%%
\subsection{$l\ge 2$ matching}
%%%%%%%%%%%%%%%%%%%%%%%%%%%%%%%%%%%%%%%%%%%%%%%%%%%%%%%%%%%%%%%%%%%%%%%%

In the interior the principal part of the free evolution equations is
$-\ddot\chi+\chi''+\dots \psi'$, $-\ddot k + c_s^2 k''+\dots\psi'$ and
$\dot\psi$. The evolution equations therefore have 5 characteristics,
namely the light cone, the sound cone, and the fluid 4-velocity. Two
of these characteristics travel from the outside to the inside, and so
two quantities can be determined freely on the stellar surface. A
convenient choice of these is $\chi$ and $k$. Similarly, the principal
part of the free evolution equation in the exterior is $-\ddot Z +
Z''$, so that we can specify one quantity freely on the stellar
surface. It is usefully chosen to be $Z$. The matching problem now
consists in determining $Z$ just outside from $\chi$, $k$ and $\psi$
and their derivatives just inside (extraction), and $\chi$ and $k$
just inside from $Z$ and its derivatives just outside (injection).

In section \ref{section:fluidstar}, we gave the continuous quantities
$C_1$ to $C_6$ in tensor components in the fluid frame.  In order to
write down equalities between quantities just inside and just outside
the stellar surface, we must express these quantities in terms of $Z$
and its derivatives. For numerical work one would probably want to use
comoving coordinates in the star and Bondi or polar-radial
coordinates outside. Therefore we use fluid frame derivatives inside,
but radial frame derivatives outside.

Frame tensor components and related quantities in any two frames are
related by hyperbolic rotations through an angle $\xi$ (for vectors)
or $2\xi$ (for 2-tensors). We use the formulas given in Appendix E of
Paper I, and take hatted quantities to refer to the radial frame and
unhatted quantities to the fluid frame. From (\ref{radial_frame}), and
Eq. (E4) of Paper I, we find that
\begin{equation}
\label{xi}
\sinh\xi=arU, \qquad
\cosh\xi=arW.  
\end{equation}
These expressions are evaluated just inside the surface.

We write down some matching conditions that we shall need as
intermediate results.  The continuous fields $C_1=N$ and $C_2=k$ are
frame-independent, and therefore they are the same just inside and
just outside the surface. The tensor components $\phi\equiv\chi+k$
and $\psi$ are also continuous, and we just need to transform then
from the fluid to the radial frame. The continuity of $C_3$ and $C_4$
therefore gives
\begin{eqnarray}
\label{phirot}
\phi & = &\cosh2\xi\ \hat\phi-\sinh2\xi\ \hat\psi, \\
\label{psirot}
\psi & = & \cosh2\xi\ \hat\psi-\sinh2\xi\ \hat\phi,
\end{eqnarray}
where the left-hand side is just inside, and the right-hand side just
outside. $k'$ and $\dot k$ transform by a rotation through the angle
$\xi$. $\dot k$ is continuous because $k$ is, but $k'$ just outside
is equal to $k'+8\pi\rho N$ just inside. Putting this together, the
continuity of $C_2$ and $C_5$ gives
\begin{eqnarray}
\label{kprimerot}
k' + 8\pi\rho N & = & \cosh\xi\ \pr{k} + \sinh\xi\ \dr{k}, \\
\label{kdotrot}
\dot k & = & \cosh\xi\ \dr{k} + \sinh\xi\ \pr{k}.
\end{eqnarray}

We obtain the extraction equation from the definition
(\ref{Zerilli_def}) of $Z$, the inverse of (\ref{phirot},\ref{psirot})
and the inverse of (\ref{kprimerot},\ref{kdotrot}). It is
\begin{equation}
\label{extraction1}
Z=A\left(\cosh2\xi\ \phi +\sinh2\xi\ \psi\right)
+(B-A)k 
+ C\left[\cosh\xi\ (k'+8\pi\rho N)-\sinh\xi\ \dot k\right],
\end{equation}
where the left-hand side is evaluated just outside and the right-hand
side just inside, and where $A$, $B$ and $C$ are the coefficients
defined in Eq. (\ref{Zerilli_coeffs}).  Substituting these
coefficients, the expressions (\ref{xi}), and the definition
$\phi=\chi+k$, we obtain the final version of our extraction equation
\begin{equation}
\label{extraction}
Z=rk+{2r^4\over(l+2)(l-1)r+6m} \left[(W^2+U^2)(\chi+k)+2UW\psi
-W(k'+8\pi\rho N)+U\dot k\right].
\end{equation}

The injection equation for $k$ is obtained from the constraint
(\ref{kZconstraint}) and continuity of $k$. The injection equation for
$\chi$ is obtained from the continuity equation (\ref{phirot}), the
definition $\phi=\chi+k$, the injection equation for $k$, and the
constraints (\ref{chiZconstraint}) and (\ref{psiZconstraint}). The two
injection equations are therefore
\begin{eqnarray}
\nonumber
k &=& (\ref{kZconstraint})
\\
\label{kinjection}
& = & {Z\over r} + {2\over l(l+1)}\left(1-{2m\over r}\right)
\left[Z_{,r}-{6m\over(l+2)(l-1)r+6m}{Z\over r}\right], \\
\nonumber
\chi &=& \cosh 2\xi \ (\ref{chiZconstraint})
- \sinh 2\xi \ (\ref{psiZconstraint})
+ (\cosh 2\xi\, - 1) \, (\ref{kZconstraint}) \\ 
\label{chiinjection}
&=& \left(1-{2m\over r}\right)^{-1} r^2 \left [
(W^2+U^2) \ (\ref{chiZconstraint})
- 2UW \ (\ref{psiZconstraint})
+ 2 U^2 \, (\ref{kZconstraint}) \right], 
\end{eqnarray}
where the equation numbers in brackets stand for the right-hand sides
of those equations, evaluated just outside the surface, and where the
left-hand sides are just inside the surface. We have not written out
(\ref{chiinjection}) in full because no important simplifications
occur when one expands it. Equations
(\ref{extraction}-\ref{chiinjection}) are the main results of this
paper.

We have not used the continuity of $C_6$.

%%%%%%%%%%%%%%%%%%%%%%%%%%%%%%%%%%%%%%%%%%%%%%%%%%%%%%%%%%%%%%%%%%%%%%%%
\subsection{$l=1$}
%%%%%%%%%%%%%%%%%%%%%%%%%%%%%%%%%%%%%%%%%%%%%%%%%%%%%%%%%%%%%%%%%%%%%%%%

For $l=1$, the polar perturbation variables we use are no longer fully
gauge-invariant. It is necessary to make a partial gauge choice, and
in Paper I we chose $k=0$. This still leaves a residual gauge freedom
worth one free function of time only. In paper I we fixed this
remaining freedom by demanding that at the center the leading order in
$r$ of the variable $\eta$ vanishes, or $\bar\eta=O(r^2)$ in the
notation of Paper I. In the exterior we also make the partial gauge
choice $k=0$, and again this leaves a residual gauge freedom
parameterized by one free function of time. As reviewed in Appendix
\ref{appendix:l=0,1exterior}, this freedom can be used to set {\it
all} perturbation variables in the exterior equal to zero, which shows
that the exterior perturbations are pure gauge. This gauge is not the
same as the one fixed by imposing $\bar\eta=O(r^2)$ at the center. We
therefore drop the latter -- it was imposed only in the absence of a
better choice and does not simplify the field equations. The equations
of Paper I are valid for any way of fixing the residual gauge freedom,
and in particular for the gauge choice $\eta=0$ at the surface that we
adopt now.

With the complete gauge choice $k=0$ everywhere and $\eta=0$ at the
surface we find that $\hat\chi=\hat\psi=\eta=0$ in the
exterior. Transforming this to to the fluid frame, we have
$\chi=\psi=\eta=0$  just outside the surface.  From the continuity of
$\tilde C_2$, $\tilde C_3$ and $\tilde C_4$, we have
\begin{eqnarray}
\label{l=1psibound}l=1,k=0,\eta=0\hbox{ just outside}: \quad \psi &=& 0, \\
\label{l=1chiibound}
\chi &=& 2\eta , \\
\label{l=1etabound}
\eta &=& {4\pi\rho \over W} N,
\end{eqnarray}
just inside the surface. These equations can be used as boundary
conditions for the integration of the ODE system (A12-A14) of Paper I.
From the continuity of $W\tilde C_5-U\dot{\tilde C_2}$, we have
\begin{equation}
\label{l=1derivbound}
r|v|^2 D(\chi-2\eta) + 8\pi\rho(\nu+W)N = 0,
\end{equation}
just inside the surface. (See Paper I for the definition of the radial
derivative $D$.) If we now use Eqs. (A13) and (A12) of paper I to
eliminate $D\eta$ and $D\chi$, (\ref{l=1psibound}-\ref{l=1etabound})
to eliminate $\chi$, $\psi$ and $\eta$, and (\ref{Nomega}) to
eliminate $N$, and use $p=0$ at the surface, we obtain an identity.

For $l=0$ and $l=1$ we use the matter perturbations $\omega$ and
$\gamma$ as dynamical variables. Together they describe sound waves,
and therefore we need one boundary condition at the surface. It is
given by (\ref{Nboundary}).

%%%%%%%%%%%%%%%%%%%%%%%%%%%%%%%%%%%%%%%%%%%%%%%%%%%%%%%%%%%%%%%%%%%%%%%%
\subsection{$l=0$}
%%%%%%%%%%%%%%%%%%%%%%%%%%%%%%%%%%%%%%%%%%%%%%%%%%%%%%%%%%%%%%%%%%%%%%%%

The situation for $l=0$ is similar to the one for $l=1$. We have
already made the partial gauge choice $k=0$.  In order to fix the
gauge further, we impose the further gauge choice $\hat\psi\equiv
-\hat n^A \hat u^A k_{AB}=0$, where $\hat\psi$ is the equivalent of
the frame component $\psi$ in the radial frame, instead of the fluid
frame. (The reason for this choice is discussed in paper I.) This
gauge choice is equivalent to
\begin{equation}
\hat\psi=0,k=0:\quad \psi={2UW\over W^2+U^2}(\eta-\chi).
\end{equation}
We shall use this to eliminate $\psi$ from $\hat C_3$ and $\hat C_4$. 

The gauge choice $k=\hat\psi=0$ still leaves a small residual gauge
freedom. In Paper I we used this last gauge freedom to fix
$\bar\eta=O(r^2)$ at the center, but as for $l=1$ this is not the best
choice to make in the presence of a vacuum exterior, where all
perturbations are pure gauge. Instead we fix the gauge so that all
perturbations vanish in the exterior (see Appendix
\ref{appendix:l=0,1exterior}), so that $k=\eta=\hat\psi=\hat\chi=0$
everywhere outside, and therefore also $\psi=\chi=0$ . Continuity of
$\hat C_2$ and $\hat C_3$ then gives rise to the boundary conditions
\begin{eqnarray}
\label{l=0chiibound}
l=0,k=0,\hat\psi=0,\eta=0\hbox{ just outside}: \quad \chi &=& 2\eta , \\
\label{l=0etabound}
\eta &=& {W^2+U^2\over W^2-U^2} {4\pi\rho \over W} N.
\end{eqnarray}
The continuity of $\hat C_4$ gives again rise to an identity, similar
to the $l=1$ case. Again, we need the matter boundary condition
(\ref{Nboundary}).

%%%%%%%%%%%%%%%%%%%%%%%%%%%%%%%%%%%%%%%%%%%%%%%%%%%%%%%%%%%%%%%%%%%%%%%%
\acknowledgments
%%%%%%%%%%%%%%%%%%%%%%%%%%%%%%%%%%%%%%%%%%%%%%%%%%%%%%%%%%%%%%%%%%%%%%%%

We would like to thank Bob Wald for helpful discussions. JMM thanks
the University of Chicago for hospitality. This research was supported
in part by NSF grant PHY-95-14726 to the University of Chicago.

%%%%%%%%%%%%%%%%%%%%%%%%%%%%%%%%%%%%%%%%%%%%%%%%%%%%%%%%%%%%%%%%%%%%%%%%
\appendix
%%%%%%%%%%%%%%%%%%%%%%%%%%%%%%%%%%%%%%%%%%%%%%%%%%%%%%%%%%%%%%%%%%%%%%%%

%%%%%%%%%%%%%%%%%%%%%%%%%%%%%%%%%%%%%%%%%%%%%%%%%%%%%%%%%%%%%%%%%%%%%%%%
\section{Derivation of the Zerilli equation}
\label{appendix:Zerilli}
%%%%%%%%%%%%%%%%%%%%%%%%%%%%%%%%%%%%%%%%%%%%%%%%%%%%%%%%%%%%%%%%%%%%%%%%

Here we only consider the case $l\ge 2$.
Recall that in the interior, $\chi$, $\psi$ and $k$ evolve
autonomously and can be considered as the true degrees of
freedom. Three constraints give the matter perturbations $\alpha$,
$\gamma$ and $\omega$ directly in terms of derivatives of $\chi$,
$\psi$ and $k$. In vacuum, however, these matter perturbations vanish,
so that the same equations become three constraints on the Cauchy data
for $\hat\chi$, $\hat\psi$ and $k$. They are
\begin{eqnarray}
\label{kdot_constraint}
0 & = & \pr{(\dr{k})} - {\hat W} \dr{{\hat\chi}}
+ \frac{l(l+1)}{2r^2} {\hat\psi}, \\
\label{k_constraint}
0 & = & -\ppr{k}  +  \frac{l(l+1)}{r^2}( {\hat\chi} + k )
- \frac{(l-1)(l+2)}{2 r^2} {\hat\chi}
+ {\hat W} \pr{\hat\chi} - 2 {\hat W} \pr{k}, \\
\label{psi_constraint}
0 & = & 
\pr{\hat\psi} + 2 {\hat \nu} {\hat\psi} + \dr{{\hat\chi}} + 2 \dr{k}.
\end{eqnarray}
The evolution equations in vacuum are
\begin{eqnarray}
\label{chi_evolution}
-\ddr{\hat\chi}+\ppr{\hat\chi} &=& 
- 2 \left[ 2 {\hat \nu}^2 - {6m\over r^3} 
    \right] ( {\hat\chi} + k )  
-  (5 {\hat \nu} - 2 {\hat W} ) \pr{\hat\chi} 
   + \frac{(l-1)(l+2)}{r^2}{\hat\chi} , \\
\label{k_evolution}
-\ddr{k} &=&   
- {\hat W}  \pr{\hat\chi}
- {\hat \nu}  \pr{k}
- \frac{4m}{r^3} ( {\hat\chi} + k )
- \frac{(l-1)(l+2)}{2 r^2} {\hat\chi}, \\
\label{psi_evolution}
-\dr{\hat\psi} &=& 2 {\hat \nu} ({\hat\chi} + k) +\pr{\hat\chi}
\end{eqnarray}
Note that in obtaining the vacuum evolution equations from the
evolution equations inside fluid matter, we have set
$c_s^2=0$. Setting $c_s^2$ to a different formal value would
correspond to adding that value times the constraint
(\ref{k_constraint}) to the evolution equation
(\ref{k_evolution}). Here we have chosen to write the evolution
equation for $k$ in the form (\ref{k_evolution}) which does not
contain $\ppr{k}$. 

We note that in vacuum $\hat\psi$ already plays a passive role, while
$\hat\chi$ already almost obeys an autonomous wave equation,
containing only $k$ as an unwanted term. This suggests looking for a
variable of the form (\ref{Zerilli_def}) that obeys a wave equation of
the form (\ref{Zerilli_eqn}) above, to be constructed from equations
(\ref{k_constraint}), (\ref{chi_evolution}) and
(\ref{k_evolution}). Note that the new variable $Z$ should not contain
$\dr{k}$ in order to avoid bringing $\hat\psi$ back into the game.

Introducing the ansatz (\ref{Zerilli_def}) in equation
(\ref{Zerilli_eqn}) and using the above equations to eliminate second
derivatives, we obtain
\begin{equation}
- \ddr{Z} + \ppr{Z} + \hat\nu \pr{Z} 
- V_0 \pr{\hat\chi} - V_1 \hat\chi - V_2 k - V_3 \pr{k} = 0 ,
\end{equation}
where the $V_i$ are certain combinations of $A,B,C$ and $a$. It is
clear that $V_0$ should vanish:
\begin{equation}
V_0=\frac{ 2 \pr{( a C + r a^2 A )} }{ r a^2 } = 0 , 
\qquad \Rightarrow \qquad 
C = - r a A + \frac{c_1}{a} ,
\end{equation}
where $c_1$ is a constant. The coefficients $V_1/A$, $V_2/B$ and
$V_3/C$ have to be equal to each other and are all equal to the
``scattering potential'' $V$. Using our result for $C$ this gives two
coupled nonlinear differential equations of second order for $A$ and
$B$. With the choice $c_1=0$ one of them is first order and can be
easily integrated (define $\Lambda_n=-n+a^2(l^2+l+1)$ for integer
$n$):
\begin{equation}
\frac{V_1}{A}-\frac{V_3}{C}= \frac{1}{r a A} 
         \pr{\left[ -\Lambda_1 A + 2B \right]} = 0 , 
\qquad \Rightarrow \qquad
B = \frac{\Lambda_1}{2} A + c_2 ,
\end{equation}
where $c_2$ is another constant. Having expressed $B$ and $C$ in terms
of $A$, the other equation gives a single differential equation for A, 
which again can be easily integrated if $c_2=0$:
\begin{equation}
\frac{V_2}{B}-\frac{V_1}{A}
=\frac{l(l+1)}{\Lambda_1} \frac{a}{A}
          \pr{\left( \Lambda_3 \frac{A}{r} \right)}
          = 0, 
\qquad \Rightarrow \qquad
A = \frac{ r c_3 }{ \Lambda_3 } .
\end{equation}
We fix the overall constant $c_3$ in the definition of $Z$ as $c_3=2$,
which makes our $Z$ equal to the variable found by Moncrief.  Using
these expressions for $A$, $B$, $C$ we finally obtain the coefficients
given above in (\ref{Zerilli_coeffs}).

The initial data $Z$, $\dr{Z}$ can be set freely on a Cauchy surface,
and $Z$ evolves autonomously. The metric perturbations $k$, $\hat\chi$
and $\hat\psi$ can be reconstructed from $Z$ using the vacuum
perturbative Einstein equations (\ref{kdot_constraint}),
(\ref{k_constraint}) and (\ref{psi_constraint}) given in the
appendix. When we introduce the expression for $Z$ into these
equations the derivative terms cancel out and $k$, $\hat\chi$ and
$\hat\psi$ are obtained as algebraic expressions in $Z$ and its
derivatives:
\begin{eqnarray}
\label{kZconstraint}
k &=& \frac{Z}{r} +\frac{2}{a
     l(l+1)}\left(\frac{6\pr{a}Z}{a\Lambda_3}+\pr{Z}\right), \\
\label{chiZconstraint}
\hat\chi &=& -\frac{2Z}{r}
      -\frac{6\pr{a}Z}{\Lambda_3}
 +\frac{2r}{l(l+1)}\left(\frac{6\pr{a}Z}{a\Lambda_3}+\pr{Z}\right)\pr{},
\\
\label{psiZconstraint}
\hat\psi &=& -\frac{2}{l(l+1)}
      \left(\frac{a(l-1)(l+2)\dr{Z}}{\Lambda_3}+r\pr{(\dr{Z})}\right).
\end{eqnarray}
Note that $\pr{a}=-a^2m/r^2$. The remaining perturbative vacuum
Einstein equations (\ref{chi_evolution}), (\ref{k_evolution}) and
(\ref{psi_evolution}) are now linear combinations of the Zerilli
equation and its derivatives. $Z$ is therefore the true degree of
freedom in the exterior.

%%%%%%%%%%%%%%%%%%%%%%%%%%%%%%%%%%%%%%%%%%%%%%%%%%%%%%%%%%%%%%%%%%%%%%%%
\section{$l=0,1$ exterior polar perturbations}
\label{appendix:l=0,1exterior}
%%%%%%%%%%%%%%%%%%%%%%%%%%%%%%%%%%%%%%%%%%%%%%%%%%%%%%%%%%%%%%%%%%%%%%%%

\subsubsection{$l=1$:}
Together with $\hat\chi$, $k$ and $\hat\psi$, now we also have the
variable $\eta$. These variables are still invariant under gauge
transformations generated by a vector field $\xi^A$ on $M^2$, but not
under gauge changes generated by $r^2\xi Y^{:a}$ with $\xi$ a scalar
field.  We impose the condition $k=0$ to eliminate this freedom. The
equations of motion are then greatly simplified:
\begin{eqnarray}
\pr{\hat\chi}&=&- \frac{2a}{r} \hat\chi , \qquad
\dr{\hat\chi} = \frac{a}{r} \hat\psi , \\
\pr{\hat\psi}&=&\frac{1-2a^2}{ar} \hat\psi , \qquad
\dr{\hat\psi} = \frac{a^2-1}{ar} ( \hat\chi-3\eta ) , \\
\pr{\eta}    &=&- \frac{\hat\chi+(-3+2a^2)\eta}{ar} .
\end{eqnarray}
They can be integrated to give
\begin{equation}
\hat\chi = \frac{g(t)}{(r-2m)^2}, \qquad
\hat\psi = \frac{r \partial_t g(t)}{(r-2m)^2}, \qquad
\eta = \frac{2mg(t)-r^3\partial^2_tg(t)}{6m(r-2m)^2}  ,
\end{equation}
where the free function of time $g(t)$ is a residual gauge freedom,
not eliminated by the condition $k=0$. We are still allowed to perform 
changes of the form $\xi(t,r)=a^2(r)f(t)$, where $f(t)$ is an arbitrary
function of time. Under that change we find $\Delta g(t)=6m f(t)$. We 
can use this freedom to choose $g(t)=0$, so that all our perturbations 
vanish. This can be interpreted as a small displacement of the 
perturbed star so that the center of mass position is not perturbed, 
and therefore the vacuum exterior is not affected by the perturbation.

\subsubsection{$l=0$:}
In this case we use the gauge $k=\hat\psi=0$. The equations are
even simpler:
\begin{equation}
\pr{\hat\chi} = - \frac{a}{r}\hat\chi, \qquad
\dr{\hat\chi} = 0 , \qquad
\pr{\eta} = 0 .
\end{equation}
Their solution is
\begin{equation}
\hat\chi = \frac{2\Delta m}{r-2m} , \qquad
\eta = g(t) , 
\end{equation}
where the constant $\Delta m$ is a perturbation of the mass of the
star and $g(t)$ is an arbitrary function of time which again
represents residual gauge freedom, now related to time
reparameterization.  To first order in perturbations, we can write the
perturbed metric as
\begin{equation}
-\left(1-\frac{2m+2\Delta m}{r}\right)\left\{[1+g(t)]dt\right\}^2
+\left(1-\frac{2m+2\Delta m}{r}\right)^{-1}dr^2
+r^2(d\theta^2+\sin^2\theta d\phi^2) .
\end{equation}
The mass perturbation $\Delta m$ is constant in time. In physical terms this
is so because the star can lose or gain mass-energy only to second
order in perturbation theory. We consider only $\Delta m=0$ here, and
treat any change in the mass of the star as a change in the
background solution.

%%%%%%%%%%%%%%%%%%%%%%%%%%%%%%%%%%%%%%%%%%%%%%%%%%%%%%%%%%%%%%%%%%%%%%%%
\section{Calculation of $N$ in terms of fluid variables}
\label{appendix:N}
%%%%%%%%%%%%%%%%%%%%%%%%%%%%%%%%%%%%%%%%%%%%%%%%%%%%%%%%%%%%%%%%%%%%%%%%

We have derived Eq. (\ref{Nomega}) in two ways. 

1. The surface of a perfect fluid star is defined by the vanishing of the
pressure $p$. We identify $f$ inside the star with
the negative fluid pressure $-p$ (the sign is chosen so that $f'>0$),
so that $N$ becomes
\begin{equation}
N = -(p^{,A} p_{,A})^{-1/2} \left(\Delta p - p^A p_{,A}\right) =
(p')^{-1}\rho(c_s^2\omega + C\sigma) \quad\hbox{just inside the surface}.
\end{equation}
(Do not confuse the pressure gradient $p_{,A}$ with the metric
perturbation $p_A$. Note also that $p^{,A} p_{,A}=p'^2$ because $p=0$
on the surface at all times, and that $p'<0$, which explains the
overall sign in the second equality above.) From Eq. (44) of paper I
we find that
\begin{equation}
p'=-\nu(\rho+p)
\end{equation}
and so we obtain Eq. (\ref{Nomega}) above. This derivation holds for
all $l$.

2. We can also obtain (\ref{Nomega}) without identifying $f$
explicitly with $-p$. The evolution equation for $k$, Eq. (88) of
Paper I, holds on both sides of the surface if one formally sets
$\rho=0$ in the exterior. It is therefore continuous. Let us
first assume $l\ge 2$. We bring known continuous quantities such as
$\ddot k$ to one side, and find that
\begin{equation}
c_s^2 (-k''+\mbox{other metric perturbations})
-\nu k' + 8\pi\rho C \sigma = \mbox{continuous}.
\end{equation}
Now we use Eq. (94) of paper I to replace the entire term in round
brackets by $8\pi\rho\omega$, and we use the continuity of $C_5$ to replace
$k'$ by $-8\pi\rho N$ plus a continuous term. (It is this step that
brings in $N$.) We obtain
\begin{equation}
8\pi\rho(c_s^2\omega+C\sigma+\nu N) = \mbox{continuous}
\end{equation}
The left-hand side vanishes identically in the exterior because
$\rho=0$, so by continuity it must vanish just inside the surface as
well, and we obtain Eq. (\ref{Nomega}). For $l=1$, Eqs. (88) and (94)
of Paper I still holds. We now have $\eta\ne 0$ and, by gauge choice,
$k=0$, but the final result is the same. In the case $l=0$, we can use
Eqs. (A18) and (A19) instead of Eq. (88) to obtain once again the same
result.

We have derived equation (\ref{Ndot}) for $\dot N$ in two different
ways, too.

1. The velocity perturbation is the Lie-derivative of the position
perturbation along fluid world lines:
\begin{equation}
\Delta u^\mu={\cal L}_u\Delta x^\mu
\end{equation}
and so we have, in any gauge, that
\begin{equation}
n_A\Delta (u^A)=n_A( u^B {\Delta x^A}_{|B}-\Delta x^B {u^A}_{|B})
=u^B(n_A\Delta x^A)_{|B}-u^B n_{A|B}\Delta x^A+\Delta x^B n_{A|B} u^A
=(n_A\Delta x^A)\dot{}-\mu (n_A\Delta x^A),
\end{equation}
where we have used Eq. (42) of Paper I to eliminate $n_{A|B}$. 
A simple calculation using Eqs. (52), (55), (21) and (42) of Paper I
gives
\begin{equation}
n_A \Delta (u^A) = \gamma + {\psi\over2} - (n_A p^A)\dot{}+\mu(n_A p^A).
\end{equation}
Eliminating $n_A \Delta u^A$ between the last two
results, and using (\ref{N_coords}) gives us (\ref{Ndot}). This
derivation holds for all $l$.

2. Alternatively, we can use the field equations directly. Assume
$l\ge 2$ for now. Using equation (93) of Paper I, extracting the
continuous terms and using the continuity of $C_5$ and $\dot C_5$, we
find that
\begin{equation}
8\pi (\rho+p)\left( \dot{N} - \mu N + \gamma + \frac{\psi}{2} \right)
=\mbox{continuous}.
\end{equation}
We use the same argument as above: this vanishes in the exterior, so
it must vanish just inside the surface, which gives us (\ref{Ndot}).
For $l=1$ and $l=0$ the derivation is similar, taking into account
$\eta\ne 0$ and $k=0$.

%%%%%%%%%%%%%%%%%%%%%%%%%%%%%%%%%%%%%%%%%%%%%%%%%%%%%%%%%%%%%%%%%%%%%%%%
\section{Perturbations of a time-dependent star: Comparison with Seidel}
%%%%%%%%%%%%%%%%%%%%%%%%%%%%%%%%%%%%%%%%%%%%%%%%%%%%%%%%%%%%%%%%%%%%%%%%

In \cite{Seidel1} Seidel considers axial perturbations of a general
time-dependent spherically symmetric star. His variable $U$ is
$-\beta$ in our notation. Our variable $\Pi$ is related to the
interior and exterior variables of Seidel as
\begin{equation}
\Pi\hbox{ (our notation)} = -{\psi\over l(l+1)R^2}\hbox{ (Seidel interior)} 
= -{\tilde\psi\over l(l+1)r^3}\hbox{ (Seidel exterior)}.
\end{equation}
The continuity conditions of Seidel agree with the continuity of $\Pi$
that we find.

In \cite{Seidel2} Seidel derives evolution and matching equations for
the $l=2$ gauge-invariant polar perturbations. He uses comoving
coordinates in the exterior and Schwarzschild coordinates in the
exterior. His exterior perturbation variables $\tilde Q_1$ and $\tilde
\psi$ are related to our variable Z, restricted to $l=2$, by
\begin{equation}
\label{Seidel_polar}
\tilde Q_1 = (1+\frac{3m}{2r})4Z, \qquad
\tilde \psi  = \sqrt{4\pi\over 5} \, 4Z.
\end{equation}
$\tilde\psi$ obeys the Zerilli equation. His variable $\tilde k_1$ is
equal to our $k$. We have already compared his interior variables with
ours in Appendix C of Paper I. 

Seidel notes that his interior variable $q_4$ is gauge-invariant when
restricted to the boundary and in any gauge where the perturbed
boundary is at the same coordinate location as the unperturbed
boundary (what we call surface gauge). He derives the matching
conditions in that gauge. We have
\begin{equation}
A^{-1}q_4 = 2 n^A p_A,
\end{equation}
with Seidel's notation on the left and ours on the right. This is of
course equal to $-2N$ in the surface gauge $\Delta f=0$ in which Seidel works,
and is therefore actually gauge-invariant, although Seidel does not
make that point. Seidel gives an evolution equation for $q_4$,
Equation (2.89), that can be cast in covariant form in our notation if
we assume that the energy density ($\rho$ in our notation,
$\eta\equiv(1+\epsilon)\rho$ in the notation of Seidel) reduces to the
rest mass density $\rho$ of Seidel when $p=0$. The rest mass density
$\rho$ of Seidel is linked to his background metric coefficient $A$
through the gauge condition $A^{-1}=4\pi r^2 \rho$. His equation
(2.89) is then exactly our equation (\ref{Ndot}).

Seidel's matching equation (2.84), giving $\tilde Q_1$ in terms of
interior variables is equivalent to our extraction equation, Equation
(\ref{extraction}). His second matching equation (2.86), which gives
$k$ in terms of his exterior variable $\tilde Q_1$ is equivalent to
our Equation (\ref{kZconstraint}). His equation (2.85) for $\tilde
Q_{1,r}$ is a linear combination of the first two.

%%%%%%%%%%%%%%%%%%%%%%%%%%%%%%%%%%%%%%%%%%%%%%%%%%%%%%%%%%%%%%%%%%%%%%%%
\section{Perturbations of Oppenheimer-Snyder: comparison with CPM}
%%%%%%%%%%%%%%%%%%%%%%%%%%%%%%%%%%%%%%%%%%%%%%%%%%%%%%%%%%%%%%%%%%%%%%%%

Cunningham, Price and Moncrief study the axial \cite{CPM1} (CPM1) and
polar \cite{CPM2} (CPM2) perturbations of spherically symmetric
homogeneous dust collapse (Oppenheimer-Snyder collapse).  The
background solution consists of a spherical segment of a dust-filled
closed Friedmann solution in its collapsing phase matched to a
Schwarzschild exterior. The interior metric, in the notation of CPM,
is
\begin{equation}
ds^2 = -d\tau^2 + R^2(\tau) \left(d{\chi}^2 +
\sin^2{\chi}\, d\Omega^2\right).
\end{equation}
Fluid elements are at constant ${\chi}$. The stellar surface
is at ${\chi}={\chi_0}$, corresponding to
$r=r_0(\tau)=\sin{\chi}_0 R(\tau)$.  These are
just a special case of comoving coordinates, and from Appendix C of
\cite{fluidpert1} we can read off that in this metric, in a fluid
frame,
\begin{equation}
\mu=U={1\over R(\tau)} {dR\over d\tau}, 
\quad \mu'=U'=0, \quad
\nu=0, \quad W={\cot{\chi}\over R(\tau)}.
\end{equation}
Here the left-hand sides are in our notation. (On the right-hand
sides, $\chi$ is a coordinate, not our perturbation variable of the
same name.)
We also have, from the equation of state $p=0$, that
\begin{equation}
m={4\pi\over 3}r^3\rho, \quad c_s^2=0.
\end{equation}

The variable $U$ of CPM1 is $-\beta$ in our notation. Our variable
$\Pi$ is related to the interior and exterior variables of CPM1 as
\begin{equation}
\Pi\hbox{ (our notation)}  
= -\frac{\pi_1}{l(l+1)R^2\sin^2\chi} \hbox{ (CPM interior)} 
= -\frac{\tilde\pi_1}{l(l+1)r^2} \hbox{ (CPM exterior)}.
\end{equation}
The continuity conditions of CPM1 agree with the continuity of $\Pi$
that we find.

We first discuss the polar perturbations of Oppenheimer-Snyder in our
notation. For homogeneous dust, the interior equations of motion
simplify greatly, to
\begin{eqnarray}
\label{chidust}
-\ddot\chi+\chi'' & = &
3\mu\dot\chi + 2 W\chi' + {(l+2)(l-1)\over r^2}\chi,
 \\
\label{kdust}
-\ddot k & = &
  U \dot\chi
+ 4 U  \dot k
- W  \chi'
+ 2 \left(W^2-\frac{1}{r^2}\right) ( \chi + k )
- \frac{(l-1)(l+2)}{2 r^2} \chi
\\
-\dot \psi & = & 2U\psi +\chi' .
\end{eqnarray}
$\chi$ therefore obeys an autonomous wave equation describing
gravitational waves that do not couple to matter perturbations. The
sound wave equation normally obeyed by $k$ has degenerated into an ODE along
matter world lines. $k$ and $\psi$ are therefore obtained by solving
ODEs after the autonomous wave equation for $\chi$ has been solved.

We now compare with CPM2. On an Oppenheimer-Snyder interior, our
perturbation variable $\chi$ is essentially the variable $Q_1$ of
\cite{CPM2}, and is related to their final variable $\psi$ as
\begin{equation}
\chi = \sqrt{\frac{5}{4\pi}} \, Q_1 ,  \qquad 
\chi = \frac{\sin\chi}{R(\tau)} \, \psi ,
\end{equation}
where again the notation on the right-hand side is that of CPM2, where
$\chi$ is a coordinate. We find that our equation (\ref{chidust}),
restricted to $l=2$, is equivalent to the autonomous wave equation
(II-41) for the variable $\psi$ of CPM2. The exterior variables $\tilde
Q_1$ and $\tilde\psi$ of CPM2, are the same as the ones used by Seidel
and given above in (\ref{Seidel_polar}). The exterior equation of CPM2
is just the Zerilli equation.

Although CPM2 can and do ignore $k$ and $\psi$ in the interior, they
need some information about their values at the surface for imposing
the matching conditions. CPM2 evolve a variable $\Delta$ along the
stellar surface that is related to our variable $k$ via
\begin{equation}
k = {1\over 3 \sin{\chi}} \Delta.
\end{equation}
Equation (B24) or (II-43c) of \cite{CPM2}, an ODE evolution equation
for $\Delta$, is our Eq. (\ref{kdust}), restricted to the boundary.
Equation (II-43a) of \cite{CPM2} agrees with our injection equation
for $k$, Eq.  (\ref{kinjection}), if one expresses $\pr{Z}$ through
$Z'$ and $\dot Z$ . CPM2 use it not as an injection equation for $k$
(as already discussed, they do not evolve $k$ at all), but as an
extraction equation for $Z$ of the form $AZ+B\dot Z+CZ'=Dk$, using the
value of $k$ on the boundary that is obtained from (II-43c).  The
injection equation for $\chi$ of CPM2 is (II-43b) of
\cite{CPM2}. Translated into our notation, is
\begin{equation}
{3\chi\over r} = \ddot Z + {4r\over 2r+3m} U\dot Z + {3m\over 2r+3m}
{m\over r^3} Z - {9m\over 2r+3m} {k\over r},
\end{equation}
where the right-hand side is evaluated just outside the surface, even
though the derivatives have been expressed in the fluid frame.
When we transform the derivatives on the right-hand side into the
radial frame, and then used the Zerilli equation to eliminate
$\ddr{Z}$, we find that this equation agrees with our injection
equation (\ref{chiinjection}).

%%%%%%%%%%%%%%%%%%%%%%%%%%%%%%%%%%%%%%%%%%%%%%%%%%%%%%%%%%%%%%%%%%%%%%%%
\section{Polar perturbations of a static star: Comparison with Thorne
{\it et al.}}
%%%%%%%%%%%%%%%%%%%%%%%%%%%%%%%%%%%%%%%%%%%%%%%%%%%%%%%%%%%%%%%%%%%%%%%%

The matching conditions for the $l\ge 2$ polar perturbations of Thorne
I \cite{ThorneI}, equation (19), are corrected in Thorne II
\cite{ThorneII}, equations (B1a-b), and we compare the corrected
version with our results. In the following the left-hand sides are
Thorne's notation and the right-hand sides are ours. We use the fact
that our gauge-invariant variables are identical with the
gauge-dependent ones in Regge-Wheeler gauge, which is the one Thorne
{\it et al.} use.

Equation (D4) of our Paper I contains two sign mistakes: the overall
signs of $\gamma+\psi/2$ and of $\alpha$ should be
reversed. Correcting this, we see that $e^{-\nu/2}W_{{\rm
T},t}=r^2(\gamma+{\psi\over 2})$. On a {\it static} background, this
in turn is the frame $\dot W_{\rm T}$ in our notation. Comparing with
our equation (\ref{Ndot}), we have $W_{\rm T}=-r^2 N$ (on a static
background only), where we have set the integration constant to zero
as it does not carry physical information. We also have
$H_0=H_2=-(\chi+k)$, $e^{(\lambda+\nu)/2}H_1=\psi$, and $K=-k$. On the
static background, $e^{-\lambda/2}f_{,r}=f'$. (Note that Thorne uses
$f'$ for $f_{,r}$, while we use it for the frame derivative along
$n^A$.) $e^\lambda$ is continuous because the Hawking mass is
continuous. Thorne {\it et al.} also choose $e^\nu$ to be continuous.
With these correspondences, we find that the five continuous
perturbation fields of Thorne II, (B1a-b), are equivalent to the five
\begin{equation}
\chi, \quad k, \quad \psi, \quad \chi', \quad k'+8\pi\rho N.
\end{equation}
These are our continuity conditions, given that $\mu$ vanishes for a
static background. 

The special case $l=1$ has been covered in Thorne V \cite{ThorneV}.
Using the relations (D3-D4) of Paper I we have re-obtained all their
evolution equations (14) and (15), with the one difference that their
(14a) should have an additional term $-S\gamma_{,r}/\gamma$ (notation
of Thorne, see there). We have verified that their regularity
conditions (20) at the center agree with our conditions (117-120) of
Paper I. The matching equations (21b-e) of Thorne V are equivalent to
our matching conditions (\ref{l=1contfirst}-\ref{l=1contlast}), if one
restricts the latter to a static background and assumes the gauge
choice $\eta=\hat\chi=\hat\psi=0$ made by Thorne V in the exterior.

%%%%%%%%%%%%%%%%%%%%%%%%%%%%%%%%%%%%%%%%%%%%%%%%%%%%%%%%%%%%%%%%%%%%%%%%

%%%%%%%%%%%%%%%%%%%%%%%%%%%%%%%%%%%%%%%%%%%%%%%%%%%%%%%%%%%%%%%%%%%%%%%%

\end{document}